\documentclass{epl}

\title{Structural lubricity: Role of dimension and symmetry}
\author{Martin H. M\"user  }
\institute{
   Department of Applied Mathematics, University of Western Ontario,
   London, Ontario, Canada N6A 5B7\\
}

\begin{document}

\maketitle

\begin{abstract}
When two chemically passivated solids are brought into contact, interfacial 
interactions between the solids compete with intrabulk elastic forces.  
The relative importance of these interactions, which are length-scale 
dependent, will be estimated using scaling arguments.
If elastic interactions dominate on all length scales, solids will
move as essentially rigid objects.
This would imply superlow kinetic friction in UHV, provided wear was absent.
The results of the scaling study depend on the symmetry of the surfaces 
and the dimensionalities of interface and solids.  
Some examples are discussed explicitly such as contacts
between disordered three-dimensional solids and
linear bearings realized from multiwall carbon nanotubes.
\end{abstract}

\section{Introduction}
Many small-scale devices cannot be miniaturized further, because friction 
and wear appear to be exceedingly large in nanoscale machines.  
This limitation seems to arise, because the surface to volume ratio is large 
in small systems and thus surface forces such as friction become relatively 
large.   
However, many theoretical 
studies~\cite{hirano90,shinjo93ss,buldum99prl,muser01prl,muser01tl}
and most recently an increasing number of 
experiments~\cite{hirano93,martin93prb,hirano97prl,crossley99,falvo99,dienwiebel03}
indicate that shear forces can be tremendously small 
between two atomically flat surfaces.  
These findings spur the hope for new avenues to reduce friction
in nano-scale applications.

Superlow friction was first suggested by 
Hirano and Shinjo~\cite{hirano90,shinjo93ss}.
Their calculations indicated that no instabilities occur when two atomically 
smooth copper solids slide past each other provided that 
they are sufficiently  misaligned.  
The absence of instabilities in such contacts implies that kinetic 
friction approaches zero at infinitely small velocities, even when thermal 
fluctuations are absent.
Hirano and Shinjo called such superlow friction ‘superlubricity’.  
This term is often considered unfortunate, because one may expect 
zero friction in the sliding state in analogy to superconductivity 
and superfluidity.  
However, due to the emission of sound waves, which also occur when 
instabilities are absent, a drag force linear in velocity will persist.  
It might therefore be more appropriate to call the effect structural 
lubricity, as the low friction arises from the structural incompatibility 
of the two contacting solids rather than from a Bose Einstein condensate 
of bosons.  
The main reason for the low friction is that lateral forces between 
two non-matching, rigid solids cancel systematically for incommensurate 
interfaces and stochastically between disordered surfaces, so that the 
average force per unit area decreases quickly with increasing 
contact area.

In principle, two solids can pin even when their intrinsic surface 
corrugations do not match.  
As suggested already by Amontons in 1699~\cite{amontons99}, 
solids can deform elastically so 
that the deformed structures  eventually interlock geometrically.  
However, elastic deformations would only produce the typically observed 
friction laws, if Prandtl's condition for (elastic) pinning was satisfied, 
namely the condition that for a given macroscopic configuration 
(i.e. a given relative displacement of the two solids), several 
non-equivalent, mechanically stable, microscopic configurations 
exist~\cite{prandtl28}.  
In that case, externally imposed sliding would lead to instabilities 
(or plucking motion) and ultimately to dissipated energy.
This mechanism is discussed in more detail elsewhere~\cite{muser03acp}.  
In the Prandtl (or Tomlinson) model, multistability and consequently
elastic hysteresis and finite kinitec friction occur when the maximum
curvature of the substrate potential exceeds the spring constant with
which the surface atom is bound to its lattice site.
Although the Prandtl Tomlinson model is only one dimensional,
it shows the importance of the competition between intrabulk elastic
restoring forces and interfacial interactions.
While there is a large literature on such competitions in the field of 
vortex motion in superconductors~\cite{blatter94}, 
no generic, simple, and quantitative analysis is known to the author, 
in which the effect of interfacial order and dimensionality $d_{\rm int}$ 
as well as the solids' dimensionality $d_{\rm obj}$ is discussed.
In the context of mechanical friction between solids, approaches are 
either computational, highly complex, and/or the issue of dimensionality 
and orientational alignment is discussed qualitatively.

The subject of this letter is a generic, quantitative analysis of how fast 
interfacial interactions and intrabulk elastic forces increase as the length 
scale (either of the object or of the description) increases.  
In this analysis, we will assume that it is legitimate to treat a 
block of length ${\cal L}$ as effectively rigid as long as the intrabulk 
stiffness (to be defined below) on that length scale is greater 
than the interfacial stiffness on the same length scale.
The reason is that Prandtl's 
multi-stability criterion cannot be satisfied in that situation.  
We will then analyze whether this assumption still holds when
the length scale is increased ever further.
If, at some point, the rigid-block assumption breaks 
down, then some fractions of the block may be able to move in a partially 
uncorrelated fashion, i.e., multistability and consequently
instabilities can occur that do not 
incorporate the motion of the whole block.  
In the latter case, the interface would not show structural lubricity.  
It has to be emphasized that the present study cannot predict with 
certainty whether an interface really loses its structural lubricity 
when the rigid block assumption breaks down. 
Our discussion would only suggest that there is the possibility for the loss 
of structural lubricity, or - in Aubry's words~\cite{aubry83,peyrard83} - 
for the possibility of the breaking of analyticity.
In the remaining part of this letter, I will present the
scaling arguments for interfacial and intrabulk interactions.
These arguments will then be applied to selected cases.

\section{Theory}

The starting point of our analysis is displayed in fig.~1(a).  
Two solids with flat interfaces are brought into contact.  
Let us assume that the interatomic interactions within each solid are 
strong (covalent, metallic, ionic bonds or a combination thereof), 
while the interactions between the solids are physical, i.e., 
relatively weak Lennard Jones-type forces between chemically 
passivated solids.  
In such a situation, one may assume that one can approximate the intrabulk 
interactions with harmonic springs of stiffness $k_{\rm bulk}^{\rm loc}$, 
while (except for small corrections due to small atomic displacements) 
the interactions of surface atoms with the opposing surface 
(called the ‘substrate) can be described by a function that has the 
same periodicity $a_{\rm s}$ as the substrate, i.e., in one dimension 
$V(x) = V_0 \cos(2\pi x/a_{\rm s})$, see Fig.~1(b).  
The maximum curvature of the substrate potential will be denoted as 
atomic-scale interfacial stiffness $k_{\rm int}^{\rm loc}$.  
It is possible to estimate the order of magnitude for the values of 
atomic-scale intrabulk and interfacial stiffness, 
$k_{\rm bulk}^{\rm loc}$ and $k_{\rm int}^{\rm loc}$, 
respectively.  
Their dimensions are given by elastic constant times interatomic 
spacing $\sigma$.  
Strong bonds lead to bulk modules of typically more than 40~GPa, 
while physical bonds are usually responsible for bulk modules in the 
order of 4~GPa.  
$\sigma$ can be roughly estimated with 2~$\AA$ for chemical bonds and
with 3~$\AA$ for physical bonds.
Based on these orders of magnitude,
one may argue that $k_{\rm bulk}^{\rm loc}\approx 40$~GPa~2~$\AA$ 
and $k_{\rm int}^{\rm loc} \approx 4$~GPa~3~$\AA$.  
It is therefore obvious in such a situation that on the atomic scale, 
each (surface) atom has exactly one well-defined mechanical equilibrium
position, once the coordinates of its intrabulk neighbours relative to the
substrate are known.  
The situation is similar to that in the Prandtl or Tomlinson model 
when the elastic coupling to the lattice site is greater than the 
maximum curvature of the substrate potential.
Thus, atomic-scale instabilities cannot occur, which, however, 
does not preclude instabilities on larger length scales.

\begin{figure}
\onefigure[width=14cm]{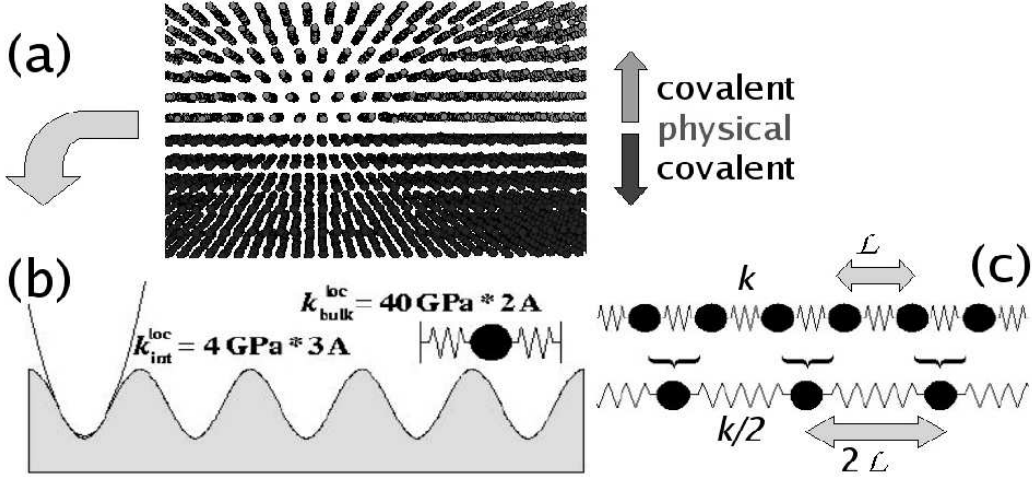}
\caption{Schematic view of the interaction between chemically 
passivated solids. 
Part (a) shows a snapshot of a simulation (whose results have been
reported in Ref.~\cite{muser01tl}) between two solids. 
Part (b) represents the coupling of a surface atom to its neighbours
(reflected by the harmonic springs) and to the substrate (reflected by
the sinusoidal line). 
The parabola indicates the maximum curvature of the atom-substrate potential,
which is called the atomic-scale interfacial stiffness $k_{\rm loc}^{\rm int}$.
Part (c) describes the scaling procedure for a 1D
elastic chain.
\label{f.1}
}
\end{figure}

In the following, we will increase the length scale of our description 
from ${\cal L} = \sigma$ to ${\cal L} = 2 \sigma$.  
Since $k_{\rm bulk}$ is much 
greater than $k_{\rm int}$ on ${\cal L} = \sigma$,  
let us group together linear blocks of length $2 \sigma$ and 
treat them as rigid objects on the new length scale.  
Of course, the values for $k_{\rm bulk}$ and $k_{\rm int}$ 
have to be renormalized for the new length scale.  
In the following ‘renormalization, we  will treat intrabulk and 
interfacial interactions completely separately.  
This is certainly a simplistic procedure as it suppresses generalized 
terms in the Hamiltonian that arise in a true renormalization 
group (RG) treatment of special case studies~\cite{joos83}. 
Thus, the present examination does not attempt to be a rigorous 
RG study.
However, it contains the leading contributions in the 
first few renormalization steps and, more importantly, it illustrates 
in an easy-to-digest fashion relevant key features for more 
detailed studies of specific systems.  
Moreover, the trends that will be obtained at the end of this analysis 
are in very good agreement with many computational and theoretical studies.
The reason for this agreement may be that the neglected additional
terms are irrelevant as long as the rigid-body hypothesis remains valid 
in the present treatment.

The coarse-graining procedure for the elastic interaction is shown in 
fig.~1(c) for a 1D elastic solid.  
Each time we group together two atoms into one new 'coarse-grained atom' by 
doubling the lattice constant in our description, we have to decrease 
the elastic stiffness of the springs by a factor of two, in order to 
maintain the same continuum limit of the chain's elastic constants.  
Thus, the effective elastic stiffness of an elastic chain on a 
length scale ${\cal L}$ (or simply the stiffness of a 
linear chain of length ${\cal L}$) is given by 
$k_{\rm bulk}({\cal L})=k_{\rm bulk}^{\rm loc}/{\cal L}$.  
If the elastic object is two dimensional, $d_{\rm obj}=2$, springs 
are coupled in series and in parallel.  
Therefore, $k_{\rm bulk}({\cal L})$ remains unaltered when the scale
is changed in 2D.  
Each additional dimension increases the effect of serial 
coupling and therefore

\begin{equation}
\label{eq:k_elastic}
k_{\rm bulk}({\cal L}) = ({\cal L}/\sigma)^{d_{\rm obj}-2} k_{\rm bulk}^{\rm loc}.
\end{equation}

As a short side comment, we note that the dimension $d_{\rm obj}=2$ 
(at which solids neither become ‘harder’ nor ‘softer’ as their 
length scales increase) is identical to the dimension above which the 
Debye Waller factor of a (topologically) perfect crystal remains finite, 
which implies that Bragg peaks can be observed for $d_{\rm obj}>2$ 
even when thermal fluctuations are present, while they are unstable 
against thermal fluctuations for $d_{\rm obj}<2$~\cite{mermin67}.

The coarse-graining procedure for the interfacial interactions 
depends on the symmetry of the surfaces.  
Let us first consider two commensurate solids.  
Each surface atom in the ‘slider’ experiences exactly the same curvature 
in the substrate potential as every other surface atom.
Therefore, if we group together ${\cal L}^{d_{\rm obj}}$ atoms into one 
coarse-grained atom, ${\cal L}^{d_{\rm int}}$ surface atoms will give a 
coherent 
contribution to the interfacial stiffness, leading to

\begin{equation}
\label{eq:k_com}
k^{\rm com}_{\rm int}({\cal L}) = 
({\cal L}/\sigma)^{d_{\rm int}} k_{\rm int}^{\rm loc},
\end{equation}

where $k^{\rm com}_{\rm int}({\cal L}) $ is the interfacial stiffness 
of commensurate solids on a length scale ${\cal L}$.

If  at least one of the two surfaces is disordered,  then the substrate 
potential that individual surface atoms experience will be random.  
The average surface curvature of interfacial stiffness has to be zero 
for flat surfaces and the variance be well defined.  
For the sake of simplicity, but without loss of generality, we 
will consider small correlation lengths.  
(Large correlation lengths lead to larger pre-factors in the following
equation, but leave the scaling unaltered.)  
If we sum up the individual (random) contributions $k_{\rm int}$,  then the 
coarse-grained interfacial stiffness $k^{\rm dis}_{\rm int}({\cal L}) $ 
will also be random with zero mean and a variance that scales – 
according to the laws of large numbers as~\cite{muser01prl}:

\begin{equation}
\label{eq:k_dis}
\left\langle \left\{k^{\rm dis}_{\rm int}({\cal L})\right\}^2 \right\rangle 
\propto ({\cal L}/\sigma)^{d_{\rm int}}
\left\langle \left( k_{\rm int}^{\rm loc}\right)^2 \right\rangle.
\end{equation}

If both surfaces are incommensurate, then the annihilation of lateral surface 
forces that individual surface atoms experience is rather systematic.  
As a consequence, we find that $k^{\rm inc}_{\rm int}({\cal L})$ does
not increase systematically with ${\cal L}$, hence

\begin{equation}
\label{eq:k_inc}
k^{\rm inc}_{\rm int}({\cal L})
 \propto k_{\rm int}^{\rm loc} .
\end{equation}

This equation might be surprising, however, it is easily verified 
exemplarily for two simple cubic lattices of identical alignment but 
different lattice constants.  
It is important to mention that the (suppressed) proportionality 
factor in the previous two equations will in general depend on the length 
scale ${\cal L}$.  
However, these prefactors will not increase systematically with 
${\cal L}$ and for each 
specific case, it will be possible to define an upper bound for the 
proportionality coefficient.  
This upper bound will mainly be based on geometric considerations such as 
relative orientation of the solids and lattice misfit.
For instance, the upper bound of the proportionality coefficient in
eq.~(\ref{eq:k_inc}) will become large close to commensurability.

As discussed in the introduction, the crucial step is to compare how quickly 
$k_{\rm int}$ and $k_{\rm bulk}$ change with the (coarse-grain) length ${\cal L}$.  
If the reduced interfacial stiffness $\kappa({\cal L})$ defined as
the ratio of interfacial and intrabulk stiffness 
\begin{equation}
\label{eq:kappa}
\kappa({\cal L}) = k_{\rm int}({\cal L})/k_{\rm bulk}({\cal L})
\end{equation}
remains smaller than unity 
for every value of ${\cal L}$, then the rigid body assumption remains valid.
The reason is that
there is a well-defined equilibrium position 
of a coarse-grained atom for every given coordinates of neighbours 
on every length scale.
Consequently, the system shows structural lubricity according to Prandtl's 
condition.  
If, in the scaling limit ${\cal L} \to \infty$ or for any intermediate 
value of ${\cal L}$, the rigid body hypothesis potentially breaks down, 
i.e., $\kappa({\cal L}) > 1$,
then the system might lose its structural lubricity.
To summarize, our analysis only addresses the question whether or not
we have to expect the solids to move as rigid units.
When they do not move as a rigid unit, our present treatment cannot
state whether motion can occur in terms of (continuous) solitons, in which
case even athermal kinetic friction would vanish at zero sliding velocity,
or in terms of (discrete) dislocations with finite kinetic friction.

\section{Examples}

{\bf\it One-dimensional, incommensurate solids.}
The Frenkel-Kontorova (FK) model, which is a 1D elastic chain embedded
in a sinusoidal external potential, is the generic model for this 
class~\cite{braun98,braun04}.
The FK model is one of the most studied models in the context of friction. 
We can easily see from eqs.~(\ref{eq:k_elastic}), (\ref{eq:k_inc}), 
and (\ref{eq:kappa}) that $\kappa$ increases linearly with ${\cal L}$
in this class.
Thus, from our discussion, we have to allow for the possibility that 
the system shows friction.
However, as emphasized before, a more detailed study has to be done 
whenever $\kappa$ exceeds unity.  
Systematically correct solutions of the FK model were provided
by Aubry and coworkers, in particular Peyrard~\cite{aubry83,peyrard83}.  
They found that for every irrational ratio $\tilde{a}=a_{\rm c}/a_{\rm s}$
of the chain's average lattice constant $a_{\rm c}$ and the substrate's lattice 
constant $a_{\rm s}$, there is a critical ratio of 
$\kappa_c = k_{\rm int}/k_{\rm bulk}$ above which analyticity is broken, 
or structural lubricity is lost.  
$\kappa_c$ is close to unity when $\tilde{a}$ equals the
golden mean and smaller for other values of $\tilde{a}$.

Nanotubes are also quasi 1D structures that can form 1D-incommensurate 
interfaces, for instance when a nanotube is placed onto graphite 
out of registry or when two nanotubes are nested into each other 
in an incommensurate fashion (i.e. in the armchair/zigzag configuration).  
Thus nanotubes do not necessarily behave structurally lubric, 
even when out of registry.  
We yet have to expect superlow friction, even structural
lubricity, because the reduced interfacial stiffness $\kappa$
at the atomic scale will be much smaller than unity.
An estimate would indicate that $k_{\rm bulk}$ is proportional to the 
stiffness of graphite's in-plane covalent bond times the diameter $R$ of 
the nanotube, while $k_{\rm int}$ would be a typical value for $k_{\rm int}$ 
based on Lenard Jones type 
interactions, hence
$\kappa(\sigma) \approx C_{44}\sigma / C_{11} R < 10^{-3}$,
where $C_{11}$ and $C_{44}$ are elastic constants of graphite.
For nested nanotubes, one may expect even smaller values of 
$\kappa_{\rm int}(\sigma)$ due to periodic-boundary conditions of the toruses.
Thus, while nanotubes will probably not move as rigid blocks,
the value of $\kappa(\sigma)$ is so small that analyticity will
most likely not be broken.





Of course, this estimate does not include effects due to the opening of
the nanotubes. Recent computer simulations indicate that
dangling bonds are present at the entrance of multiwall nanotubes.
These bonds are chemically active and explain discrepancies between
theoretically predicted and experimentally measured shear 
forces~\cite{stuart04}.

{\bf\it Three-dimensional disordered solids with two-dimensional interface.}
A generic case of solid sliding is that between two 3D solids with a 
2D interface.  
If the surface is disordered, then the interfacial stiffness increases 
as quickly with length scale as the intrabulk elastic forces.  
In situations, where restoring forces show the same scaling as random forces, 
the fluctuations typically win on exponentially large length scales.  
(This phenomenon is well-known when studying elastic neutron scattering 
of 2D solids at non-zero temperature, where the intensities of 
Bragg reflexes decrease logarithmically with increasing system 
size~\cite{mermin67}.)  
Indeed, Persson~\cite{persson99} and 
Sokollof~\cite{sokollof02} found such logarithmic laws.  
They studied the elastic coherence length $\lambda$
(also called Larkin length~\cite{blatter94}, the length scale 
below which a block is believed to move as a highly correlated unit and 
above which it moves less correlated), and 
found that $\lambda$ depends exponentially on elastic constants and other 
parameters.  
Thus, the present scaling study is in qualitative agreement with 
their results for sufficiently flat, disordered interfaces. 
If, however, the surfaces are fractal, which is the case for almost any
macroscopic object, then the random lateral forces will
cancel less quickly with increasing coarse-graining than
for flat surfaces.
It may therefore be more realistic to expect a powerlaw dependence of 
$\lambda$ on elastic constants and atomic-scale 
interfacial stiffnesses.  
Yet, the net friction force that comes from
pinning due to surface corrugation would remain extremely small.

{\bf\it Three-dimensional, incommensurate crystals.}
The ratio of interfacial and intrabulk stiffnesses becomes very small for 
3D crystals that form an atomically smooth, incommensurate 
interface, i.e., $\kappa$ decreases with $1/\sqrt{L}$.  
Therefore, if plastic flow and surface contaminations are absent,
we should expect structural lubricity, unless the orientation 
of the like solids is close to commensurability.  
Clean, incommensurate interfaces have been studied exceedingly by computer 
simulation and essentially all studies based on interatomic potentials
point to structural lubricity, in some cases even for non-passivated
surfaces, i.e., the original studies on 
superlubricity~\cite{hirano90,shinjo93ss}.
A more detailed overview is given in Ref.~\cite{muser03acp}.

\section{Conclusions and Outlook}

In this letter, I discussed the competition between interfacial interactions
and intrabulk elastic forces when two chemically passivated solids
are brought into mechanical contact.
Simple scaling arguments were provided that allow one to estimate
whether structural lubricity (essentially zero kinetic friction in the 
small-velocity limit) should be expected in UHV 
(as long as plastic deformation does not occur).
Some specific examples were discussed in terms of the scaling arguments.
For instance, the analysis showed that embedded nanotubes are not 
structurally lubric in a strict sense, even though interfacial
interactions are too weak to break analyticity.
Two incommensurate 3D crystals forming a flat 2D interface
should almost always be structurally lubric.
However, if the surfaces are disordered and moreover fractal, then
structural lubricity will most likely be lost. 
The latter result would imply that different asperities in a typical
multi-asperity contact do not move in a correlated way.

The present manuscript focused on flat or at least self-similar surfaces.
A separate treatment would have to be done for curved surfaces.
First steps in that direction have been taken in a computer simulation study 
of an AFM tip that is pressed against a substrate~\cite{wenning01}.
Another interesting issue would be to study the effect of thermal
and quantum fluctuations on the drag forces between structurally
lubric solids.
(Most studies so far have been done for the 1D FK model, which
probably does not show the same universal behaviour as 3D solids, 
as the reduced interfacial stiffness tends to 
infinity at large length scales for the 1D FK model.)
Studying the (quantum) friction between two incommensurate He$^4$ solids 
is certainly computationally feasible.
It may be speculated that true superlubricity occurs below the temperature 
at which solid He$^4$ condenses into a 'supercrystal'.

It will certainly remain an experimental challenge to prepare sufficiently 
flat, clean, and chemically inert surfaces to make technological
use of structural lubricity. 
Moreover, structural lubricity will break down at normal pressures 
that are high enough to deform solids plastically.
At least this latter point might not be an issue in nano-scale devices,
as they are less rough than normal macroscopic surfaces and therefore
have less extreme pressure distributions.



\acknowledgments
This research was funded by the Natural Sciences and 
Engineering Research Council of Canada (NSERC) and
Sharcnet.

\end{document}